\documentclass[prd,twocolumn,floatfix,preprintnumbers,letterpaper]{revtex4}
\usepackage{graphics}
\usepackage{epsfig}

\usepackage{natbib}
\usepackage{amsmath}

\newcommand{\loga}{\log(a)}
\newcommand{\Cl}{\mathcal{C}_{\ell}}
\newcommand{\zrec}{z_{\text{\it rec}}}
\newcommand{\percent}{\%}

\begin{document}

\title{Constraints on Non-Standard Recombination from Recent CMB Observations}
\author{Angela M. Linn}
\affiliation{Department of Physics, The Ohio State University,
Columbus, OH 43210}
\email{linn.24@osu.edu}
\author{Robert J. Scherrer}
\affiliation{Department of Physics and Astronomy, Vanderbilt University, Nashville, TN 37235}
\email{robert.scherrer@vanderbilt.edu}
%
\begin{abstract}  We examine cosmic microwave background constraints on variations in the recombination history
of the universe.
We use a very general extension to the standard model of recombination 
characterized by two parameters, $a$ and $b$, which multiply the overall rates of recombination and 
ionization and the binding energies of hydrogen respectively.  We utilize WMAP temperature and TE cross-correlation data, 
along with temperature data at smaller scales from ACBAR and CBI, to place constraints 
on these parameters.  The range of 
recombination histories which gives the best fit to the data is
$-0.6 < \loga < 0.5$ and $0.85 < b < 1.15$, forming a diagonal region from 
low-$a$/high-$b$ to high-$a$/low-$b$.  We find $z_{rec} = 1055 \pm 25$ and $\Delta z = 198 \pm
5$ at $68\percent$ confidence 
for the range of recombination models which are best able to reproduce the observed CMB power 
spectra.  Standard recombination provides an acceptable fit to the data; while
there is still room for non-standard recombination, the allowed variation is small. 
\end{abstract}
\maketitle
%
\section{Introduction}  
\label{sec:Introduction}

The cosmic microwave background (CMB) provides a snapshot of the early universe.  The CMB radiation last interacted directly (i.e.,
non-gravitationally) with matter during the epoch of recombination, the period when the universe became cool enough for atomic nuclei to
capture and hold electrons for the first time.  Therefore, the CMB provides a sensitive test to any variations in the standard model of
recombination.
 
A number
of authors have looked at CMB constraints on variations in standard
recombination \cite{Peebles:2000, Landau:2001, DoroshkevichNaselsky:2002, Bean:2003}. 
Here we reexamine this question in the light of recent, high-quality data from the WMAP 
satellite \citep{Bennett:2003}, ACBAR \citep{Pearson:2002}, and CBI \citep{Kuo:2002}, following 
the parametrization presented in Ref. \cite{Hannestad:2000}.  The WMAP data represents
a significant improvement over data
used in most previous calculations, and significantly tightens the 
resulting constraints.  In addition, we include the analysis of polarization data in 
the search for constraints on recombination.  The only previous analysis to use the recent 
WMAP data has been given by Bean, et al. \cite{Bean:2003}, who examined changes in the ionization history
induced by Lyman-$\alpha$ photons and ionizing photons.  Our approach is somewhat different; we have
attempted a more general, model-independent investigation, based on the discussion in
Ref. \cite{Hannestad:2000}.  Hence, the constraints we derive are not tied to any
particular model for altering recombination, but are potentially more general.  Further, we have used additional data from ACBAR and CBI.
In addition to using the TE cross-correlation data, we investigate the ability of the E-mode polarization
power spectrum to further constrain recombination.    
                     
In Section \ref{sec:Methods} we describe a parameterization of the epoch of recombination and 
the details of how we obtain constraints on the parameters. In Section \ref{sec:Results} we 
display the resulting constraints on recombination and discuss how the polarization power 
spectrum is affected by recombination. In Section \ref{sec:Conclusions} we present our conclusions.

\section{Methods}  
\label{sec:Methods}
%
\subsection{Recombination Model}
\label{sec:RecombinationModel}

The ionization fraction $x_e$ is the fraction of ionized hydrogen present in the universe,
\begin{equation}
x_e = n_e/n,
\end{equation}
where $n_e$ is the number density of free electrons and $n$ is the number density of all hydrogen atoms, 
both ionized and neutral.
The evolution of $x_e$ is given by the ionization equation and describes the progress of recombination.
The evolution of $x_e$ is given by \cite{Peebles:1968}
   \begin{equation}
      \label{eqn:IonizationFraction}
      - \frac{d x_e}{d t} = C\left[R n_p x_e^2 - \beta (1-x_e)\exp\left(-
      \frac{B_1-B_2}{kT}\right)\right].
   \end{equation}
Here $C$ is Peebles' correction factor, $R$ is the recombination coefficient, $\beta$ is the ionization coefficient, $B_1$ and $B_2$ are the binding energies if the first two levels of hydrogen, and $n_p$ is the number density of free protons plus hydrogen atoms.
 
It is our goal to constrain recombination in a general 
way, rather than investigating specific mechanisms.  
To that end, we follow the framework presented 
in Ref. \cite{Hannestad:2000} for modeling a general change
in the evolution of the ionization fraction as a function of time. 
We modify the ionization history by 
inserting two parameters, $a$ and $b$, 
into Equation (\ref{eqn:IonizationFraction}):
  \begin{equation}
  \label{eqn:IonizationFractionHS}
   - \frac{dx_e}{dt} = aC_P \left[ \alpha n x_e^2  - \beta (1-x_e) e^{-b(B_1 -
B_2)/k_B T} \right].
  \end{equation}

The constant $a$ multiplies the overall rates of both recombination and ionization,
while the constant $b$ multiplies the binding energies of hydrogen.  The 
values $a=1$ (or $\loga=0$) and $b=1$ are the case of standard recombination.  We neglect 
any corresponding change in helium recombination, as that effect is small and for most 
models helium recombination is finished by the time hydrogen recombination begins.  Very 
roughly, a change in $a$ alone will change the duration of recombination, while 
keeping the onset of recombination fixed, while altering $b$ shifts the onset of 
recombination to earlier or later redshifts.
 
\subsection{Data Sets and Priors}
\label{sec:DataSetsAndPriors}

Our goal is to find a range of recombination histories --- that is, regions in
\hbox{$a$-$b$} parameter space --- which are capable of predicting CMB power 
spectra compatible with the observed CMB power spectra.
We create a grid in \hbox{$a$-$b$}, and
for each point in this space, we allow a set of five cosmological 
parameters to vary, as well as the overall normalization of the power 
spectrum.  The vector of free parameters that we vary is 
$\vartheta$=\{$\Omega_m$, $\omega_b$, $h$, $n_s$, $\tau_{RI}$, $Q$\}:  the matter density relative to critical $\Omega_m$, 
the 
baryon matter density $\omega_b \equiv \Omega_{b}h^2$, the Hubble 
parameter $h$ (in units of 100 km sec$^{-1}$ Mpc$^{-1}$),  the spectral 
tilt $n_s$, the optical depth to the surface of last scattering 
$\tau_{RI}$, and the overall normalization of the power spectrum $Q$.

We assume priors for several of these cosmological parameters.
For the Hubble parameter, 
we impose the constraint $h=0.72 \pm 0.08$, consistent 
with estimates obtained from the Hubble Space Telescope 
(HST) Key Project and observations of Type Ia supernovae 
(SNe Ia) \citep{Freedman:2001, Gibson:2000}.  The baryon 
density is taken from BBN constraints 
\citep{Steigman:2002}:  $\omega_b \equiv \Omega_b h^2 = 0.020 \pm 0.002$.  Our 
estimate for the total matter density comes from supernovae studies 
\citep{Perlmutter:1998, Riess:1998} and x-ray observations of the 
gas mass fraction of relatively relaxed galaxies \citep{Allen:2002}.  
We combine these, using generous error bars, to get $\Omega_m = 0.29 \pm 0.10$.  
These are all taken to be Gaussian, with uncertainties at the $68\percent$ confidence level.
For the spectral tilt $n_s$ we impose a uniform
prior of $0.7 < n_s < 1.3$.  Values of $n_s$ beyond 
this range are disallowed.  The reason for this is that $n_s$ 
is generally not well known; indeed, the best estimates for the 
value of $n_s$ come from the CMB itself \citep{Jaffe:2003, Spergel:2003, Benoit:2002}.  
Thus we use generous bounds which are theoretically motivated.  We do 
not include priors for the optical depth $\tau_{RI}$ or the 
overall normalization of the power spectra $Q$.  We additionally 
restrict our models to a flat $\Lambda$CDM universe ($\Omega_{tot}=1$) 
and adiabatic initial density perturbations.                  %

We then use \texttt{CMBFAST} \citep{Seljak:1996} to produce 
a CMB anisotropy spectrum from our model, as a function of 
$a$, $b$, and $\vartheta$.  For each point in \hbox{$a$-$b$} 
space we vary $\vartheta$ to minimize $\chi^2$, giving the combination 
of free parameters which is best able to reproduce the 
observed CMB power spectra for a given $a$ and $b$.

For our data we use a combined set consisting 
of the WMAP temperature \citep{Hinshaw:2003} and 
TE cross-correlation \citep{Kogut:2003} data, along 
with smaller scale temperature data from ACBAR \citep{Kuo:2002} and 
CBI \citep{Mason:2002, Pearson:2002}.  We follow the lead of Ref. \cite{Verde:2003} 
in combining the WMAP temperature data with the higher $\ell$ data from ACBAR and CBI.  A calibration uncertainty of $20\percent$ is assumed for ACBAR and $10\percent$ for CBI.   Although ACBAR has a $3\percent$ beam width uncertainty, this effect is extremely small compared to the calibration uncertainty and we neglect it. We minimize $\chi^2$ using the combination of temperature and TE power spectra.

\section{Results}  
\label{sec:Results}

\subsection{Effects of Varying the Recombination History}
\label{sec:Effects}
Altering $a$ or $b$ from their standard values has a direct effect on how recombination proceeds.  The 
easiest way to see this is by looking at the ionization fraction $x_e$ as a function of time.  
The ionization fraction determines the optical depth, $\tau$ due to Thomson scattering,
   \begin{equation}
      \tau = - \int^z_0 c\sigma_T n_e(z) (dt/dz)dz,
   \end{equation}
which in turn gives the visibility function, $g(z)$:
   \begin{equation}
      g(z) = e^{-\tau} d\tau/dz,
   \end{equation}
%

\begin{figure}[htbp]
\centering
\includegraphics[clip=]{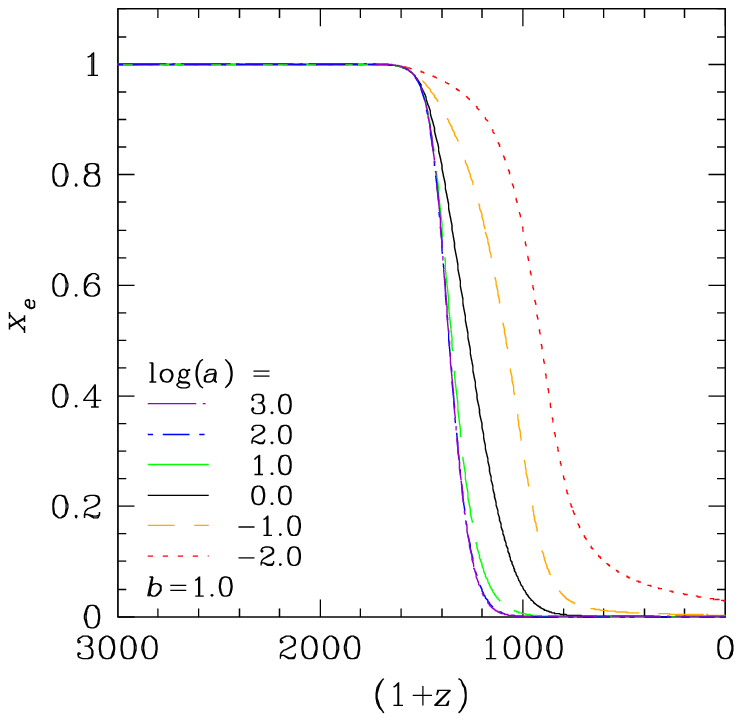}
\includegraphics[clip=]{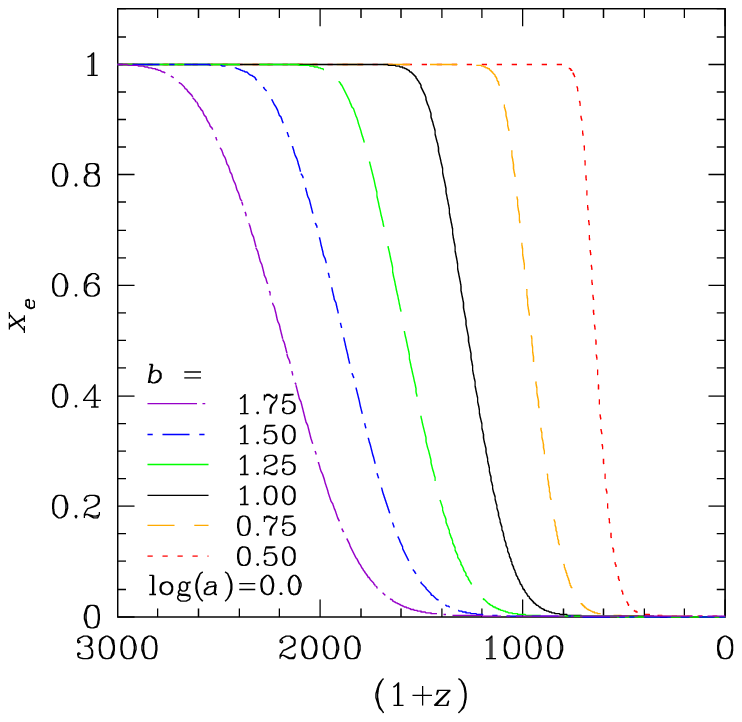}
\caption{\small The ionization fraction, $x_e(z)$, as a function of redshift, $z$, for varying $a$, fixed
$b$ (top) and for varying $b$, fixed $a$ (bottom), for $\Omega_m = 0.29$, $\omega_b=0.020$, $h=0.72$, $n_s=1.0$, $\tau_{RI}=0.15$.}
\label{fig:xe_vary_ab}
\end{figure} 
%
\begin{figure}[htbp]
\centering
\includegraphics{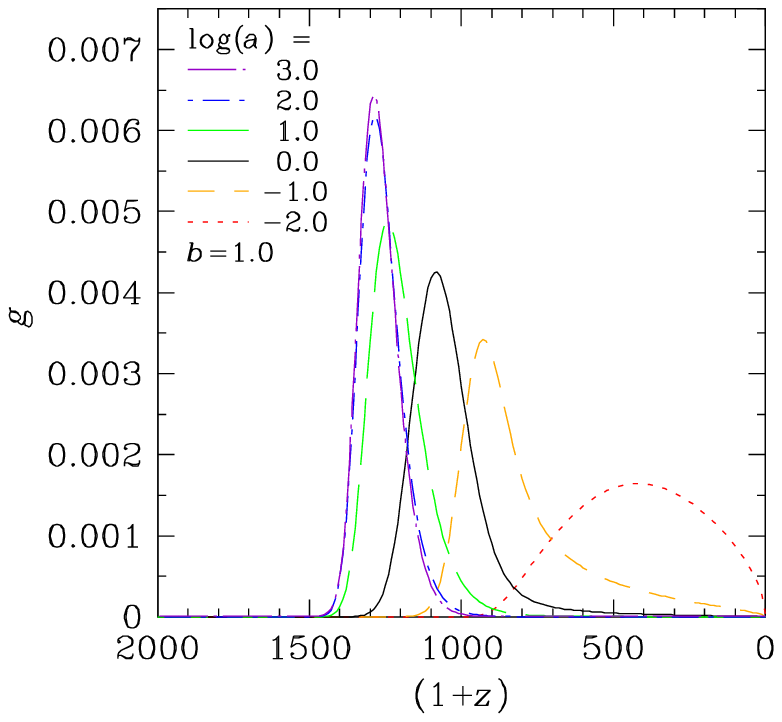}
\includegraphics{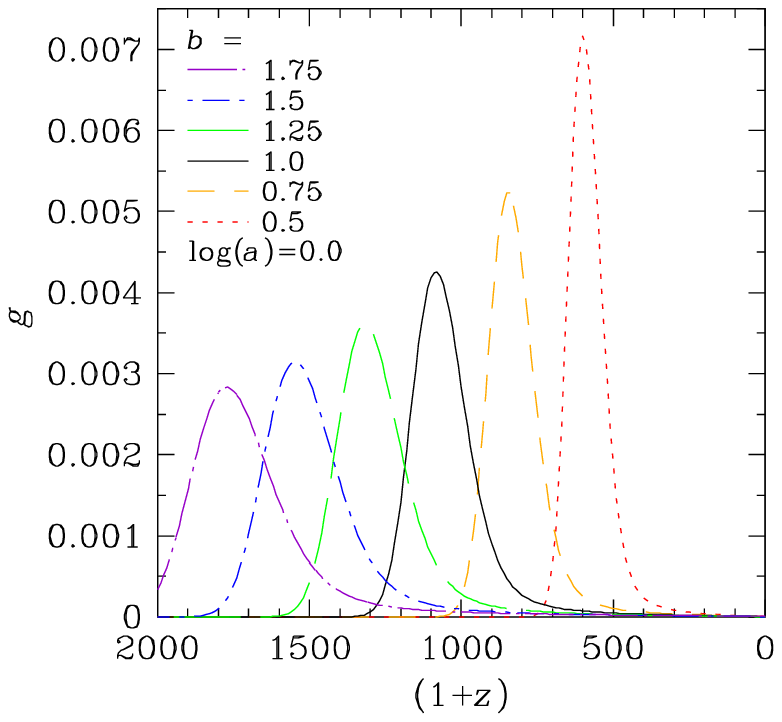}
\caption{\small As Fig. 1, for the visibility function $g(z)$.}
\label{fig:g_vary_ab}
\end{figure}

Figure \ref{fig:xe_vary_ab} shows the ionization fraction versus redshift for a variety of recombination
histories ($a$, $b$).  In each case, the free parameters $\vartheta$ are fixed at the
values \{$\Omega_m = 0.29$, $\omega_b=0.020$, $h=0.72$, $n_s=1.0$, $\tau=0.15$\}.  As $a$ increases, recombination has a longer duration, though it starts at about the same time.  Beyond a certain limit, roughly $\loga =1.5$, recombination proceeds as rapidly as possible, and increasing $a$ has no further effect because in the limit of high $a$, recombination proceeds in equilibrium.  As $b$ increases, the primary effect is that the onset of recombination shifts to earlier epochs.  A secondary effect is that the duration of recombination increases with increasing $b$.

The resulting visibility functions of the models 
used in Figure \ref{fig:xe_vary_ab} are shown in 
Figure \ref{fig:g_vary_ab}.  As $a$ decreases, 
the redshift of recombination, $\zrec$, 
moves to more recent times, and the width $\Delta z$ 
of the last scattering surface (LSS) increases.  (We define $\zrec$ to be
the redshift at which the visibility function is peaked, while $\Delta z$
is the full-width half-maximum of the visibility function).  
The degeneracy of models with sufficiently high 
values of $a$ is clearly seen. As $b$ decreases, 
the peak of the visibility function and $\zrec$ 
move toward lower redshifts, just as with decreasing $a$.  
However, with decreasing $b$, unlike decreasing $a$, the width $\Delta z$ decreases.  Thus, at the expense of allowing 
$\Delta z$ to change, changing $a$ and $b$ in opposite 
directions allows $\zrec$ to remain relatively unchanged.

Changes to the recombination history have a direct effect on the temperature and polarization anisotropy
spectra of the CMB.  Decreasing $a$ effectively increases the width of the last scattering surface
$\Delta z$, and to a lesser extent also decreases $\zrec$.  The increase in $\Delta z$ leads to
increased diffusion damping of the anisotropies at large scales so that the peaks in the
temperature, polarization, and TE cross-correlation power spectra at large scales are all less pronounced.
The decrease in $\zrec$ shifts the features of 
all three power spectra to larger scales.  

Decreasing $b$ primarily decreases $\zrec$ and also leads to a slight decrease in the value of $\Delta z$.  The decrease in $\zrec$ will again shift the features of the power spectra to larger scales.  In this case the relative amplitudes of the peaks vary considerably, whereas when $a$ is varied, the relative amplitudes remain fairly constant:  as $b$ is increased, the height of the even peaks relative to the odd peaks is also increased.  This is due to the change in the ratio of $\rho_b$ to $\rho_\gamma$ caused by later recombination.  Because $\Delta z$ increases only slightly as $b$ is increased, diffusion damping at small scales for higher values of $b$ is less significant.

\subsection{Allowed Region in \hbox{$a$-$b$} Space}
\label{sec:AllowedRegion}

\begin{figure}[htbp]
\centering
\includegraphics{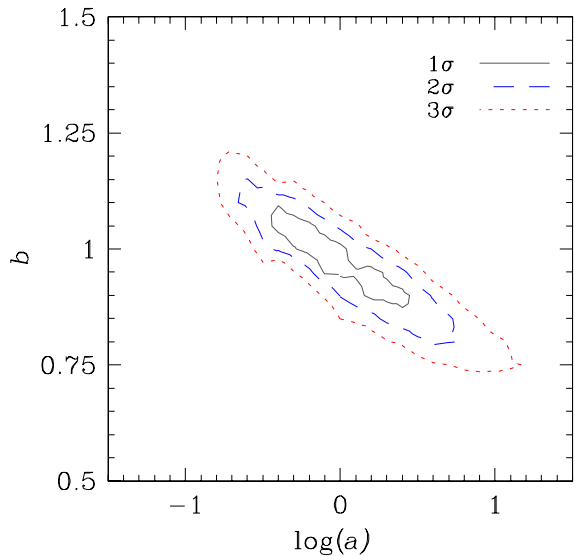}
\caption{\small Allowed region of \hbox{$a$-$b$} space.  Solid, long-dashed, and short-dashed
curves enclose the $1\sigma$, $2\sigma$, and $3\sigma$ regions.}
\label{fig:abcontour_WMAP+_TTTE}
\end{figure}

The region in \hbox{$a$-$b$} parameter space 
which is consistant with the CMB observations 
is displayed in Figure \ref{fig:abcontour_WMAP+_TTTE}.
This is the main result of our paper.
The single best fit value is $\loga=-0.2$, $b=1.0$, and standard recombination lies well inside
the $1\sigma$ region.
The allowed region of \hbox{$a$-$b$} space is considerably reduced compared to
constraints using pre-WMAP data \cite{Hannestad:2000}.  
The $3\sigma$ region extends only from $-0.8 < \loga < 1.2$ and
$0.70 \lesssim b \lesssim 1.25$.  
The $1\sigma$ regions are $-0.6 < \loga \lesssim 0.5$ and $0.85 < b < 1.15$. 

\subsection{Characteristics of the Allowed Regions}
\label{sec:Characteristics}

\begin{figure}[htbp]
\centering
\includegraphics{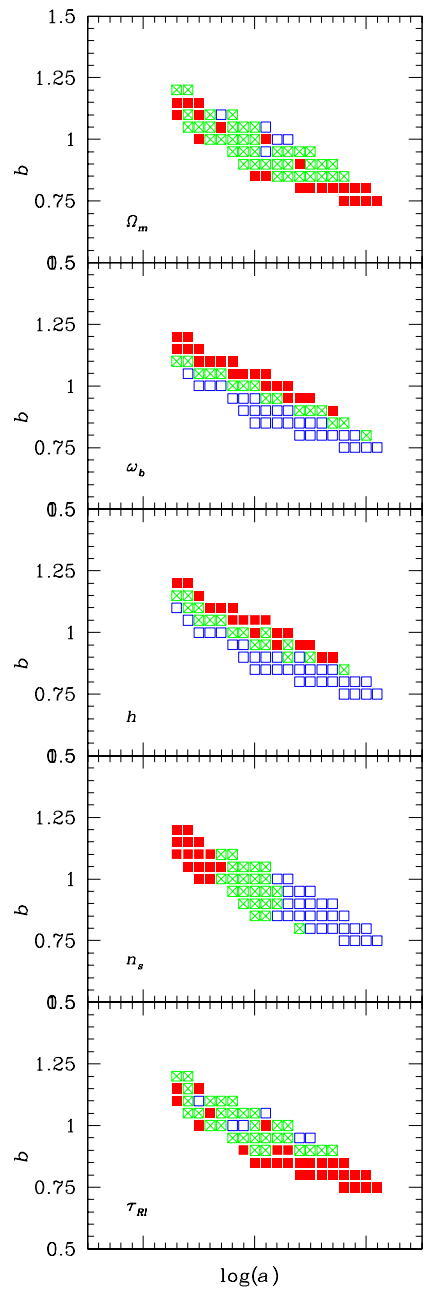}
\caption{\small How the best-fit cosmological parameters change with $a$ and $b$ for 
recombination histories falling inside the $3\sigma$ region.  The squares with an
$\times$ have ``average" values for a given parameter, the solid squares have ``high" values,
and the open squares have ``low" values.  (See text for the precise definition).}
\label{fig:paramgrid_WMAP+_TTTE}
\end{figure}

It is interesting to examine the manner in which the various cosmological
parameters change in order to compensate for changes in $a$ and $b$.
 
The shape of the CMB power spectrum is sensitive to changes in each of the cosmological parameters
\citep{Bersanelli:2002}.  Increasing the baryon density suppresses the even peaks with respect to the odd peaks because
the increased baryon mass contributes more gravity to the potential wells, lessening the effect of the rarefaction phase
of the acoustic oscillations.  Increasing the Hubble constant will lead to matter-radiation equality happening earlier
and an increased expansion rate, thus the power spectrum will tend to have suppressed even peaks and move to slightly
lower values of $\ell$.  An increase in the total matter density keeping the total density fixed at $\Omega_\text{\it
tot}=1$, or equivalently a decrease in the dark energy density, again moves matter-radiation equality to earlier times,
which tends to move the spectra to lower $\ell$ and increase the late ISW effect.
An increase in the scalar spectral index  gives more power to smaller scales, 
and will tilt the CMB power spectrum, increasing the amplitude at large $\ell$ compared to small $\ell$.
Finally, increasing the optical depth due to reionization will decrease the power at small scales by a factor $e^{-2\tau_{RI}}$.                    

Figure \ref{fig:paramgrid_WMAP+_TTTE} shows how the best-fit cosmological parameters change for 
all points falling inside the $3\sigma$ boundary.  In producing this figure, we first determine the
mean and $1\sigma$ variation for these parameters within the $1\sigma$ region in $a-b$ space.
Then we calculate the best-fit value of each parameter for a
given value of $a$ and $b$.  The
open squares with $\times$'s show points for which
the best-fit value for a given cosmological parameter lies within $\pm 1\sigma$ of the mean value.
The solid squares show
points for which the best-fit value of the given parameter is higher than this (i.e., more
than $1\sigma$ above the mean), and the open squares show points for
which the best-fit value is lower than $1\sigma$ below the mean. 

The baryon density $\omega_b$ shows a strong tendency to increase both as $a$ is increased and as $b$ is increased.  Similarly, the Hubble constant $h$ increases both with increasing $a$ and increasing $b$.  The scalar spectral index $n_s$ increases with decreasing $a$ but there does not appear to be any correlation of $n_s$ with changing $b$.  The optical depth $\tau_{RI}$ appears to increase with decreasing $b$, but this effect is not nearly so monotonic as the correlations in the previous three parameters.  There does not appear to be any significant correlation of $\tau_{RI}$ with changes in $a$.  For $\Omega_m$, there does not appear to be much correlation with changes in $a$ or $b$.
 
To summarize these effects,
an increase in $a$ is compensated by corresponding increases in $\omega_b$ and $h$ 
and a corresponding decrease in $n_s$.  An increase in $b$ is compensated by increases in $\omega_b$ and $h$ and possibly a decrease in $\tau_{RI}$.
 
The corresponding changes in the cosmological parameters as $a$ and $b$ are changed are due to competing effects.  As $a$ is increased, the peaks at high $\ell$ are less damped, because the width of recombination is decreased, and the peaks are shifted to higher $\ell$ because recombination is moved to higher redshift.  The decrease in $n_s$ lowers the amplitude of the spectrum at small scales, counteracting the increased power at small scales due to a narrower surface of last scattering.  An increase in $h$ tends to move the peaks to larger scales, counteracting the effect of a higher $\zrec$.  Similarly, increasing $b$ will enhance the height
of the even peaks by increasing $\zrec$, thus lowering the ratio of the baryon to photon density.  This is compensated by a higher values of $h$ and $\omega_b$, which tend to suppress the height of the even peaks.   The increased $h$ will again move the peaks to slightly larger scales, even as an increase in
$b$ also increases $\zrec$ and shifts the spectrum to smaller scales.

\begin{figure}[htbp]
\centering
\includegraphics{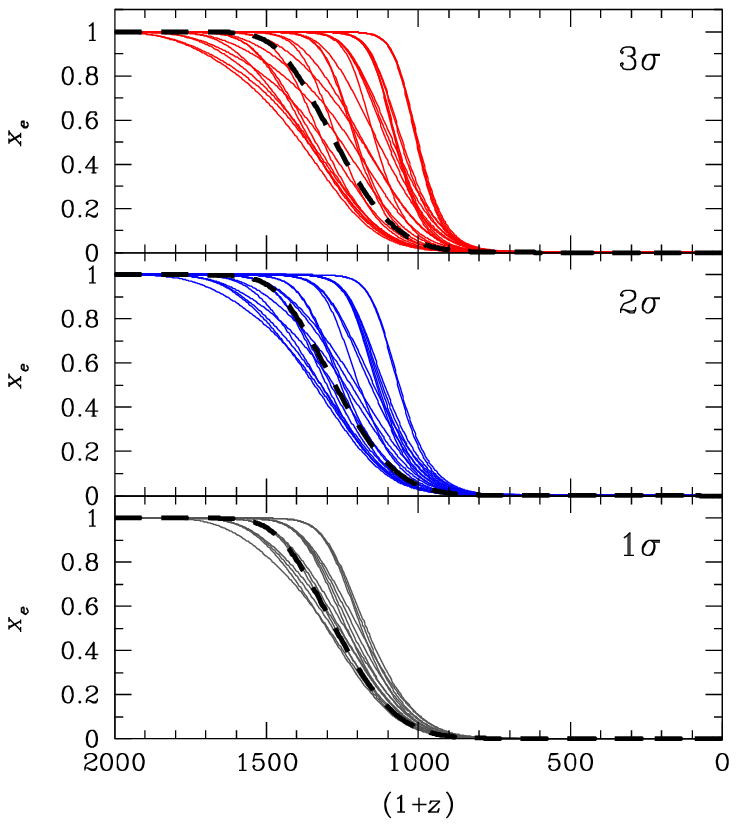}
\includegraphics{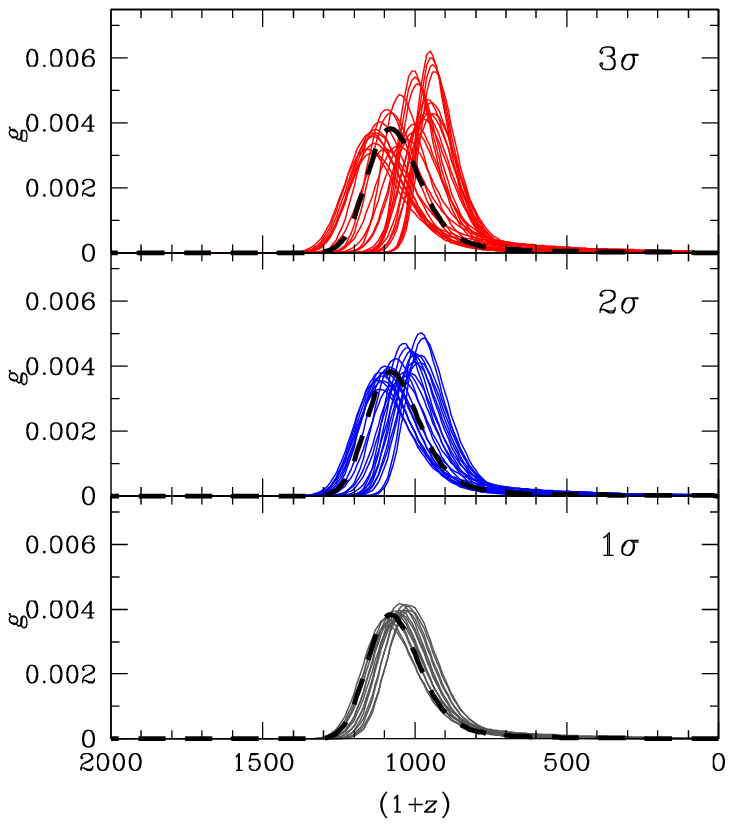}
\caption{\small Ionization fraction $x_e(z)$ vs. redshift $z$ (top) and visibility function $g(z)$ vs. redshift
$z$ (bottom) for
recombination histories falling within the $1\sigma$, $2\sigma$, and $3\sigma$
regions of the analysis.  The thicker dashed line is standard recombination, and  it falls within the $1\sigma$ region.  It is shown on the other two graphs for comparison purposes.}
\label{fig:xe_g_WMAP+_TTTE}
\end{figure}
It is also instructive to look at the evolution of the ionization fraction $x_e$ as a function of
the redshift $z$ for those recombination
histories which provide a good fit to the data. Figure \ref{fig:xe_g_WMAP+_TTTE}
shows $x_e$ versus redshift for a selection of recombination histories in the $1\sigma$, $2\sigma$,
and $3\sigma$ regions. In each case, the cosmological parameters used in calculating $x_e(z)$ were those
which produced the best fit to the CMB data.
 
From Figure \ref{fig:xe_g_WMAP+_TTTE}, it can be seen that the common feature in all recombination histories capable of producing the observed CMB is their ending point.  The recombination histories start at different times and have different durations, but all end at approximately the same era.  For the $1\sigma$ region, the average redshift at which the ionization fraction falls to $1\percent$ is $z(x_e=0.01) = 856 \pm  20$.   By contrast, the point where the ionization fraction falls to $50\percent$ has a much larger standard deviation:  $z(x_e=0.50) = 1237 \pm 44$.  All error estimates are at the 68$\percent$ confidence level.
 
We find a similar but smaller effect looking at the visibility function.  Figure \ref{fig:xe_g_WMAP+_TTTE} shows the corresponding visibility functions $g(z)$ for the $1\sigma$, $2\sigma$ and $3\sigma$ regions.
For the $1\sigma$ region, we find that the average redshift of the peak of the
visibility function is $\zrec = 1055 \pm 25$, while the average width is $\Delta z = 198 \pm 5$.  If we look at the point where the visibility function falls to $10\percent$ of its peak value, we again find a somewhat tighter constraint than at the 
peak, $z(g/g_{max}=0.10)= 811 \pm 19$, though the difference is not quite as great as that found using $x_e$.
 
The shapes of the visibility functions stray further and further from that of standard recombination in the $2\sigma$ and $3\sigma$ regions.  Although some curves in the $3\sigma$ region have a similar $\zrec$ to standard recombination, these curves tend to have a significantly different $\Delta z$.  This suggests that we are not placing a constraint on $\zrec$ or $\Delta z$ individually, but jointly.                               

\begin{figure}[htbp]
\centering
\includegraphics{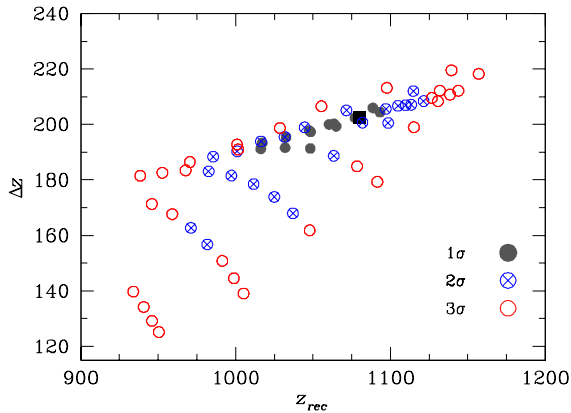}
\caption{\small Redshift of recombination $\zrec$ vs. width of the last scattering surface $\Delta z$ for recombination histories falling within the $1\sigma$, $2\sigma$, and $3\sigma$ regions of the WMAP+ TT+TE analysis.  The apparent lines in this graph are due to the finite grid spacing in $a$ and $b$.}
\label{fig:zgrid_WMAP+_TTTE}
\end{figure}

In Figure \ref{fig:zgrid_WMAP+_TTTE}, we have taken points from the $1\sigma$,
$2\sigma$, and $3\sigma$ regions of the allowed \hbox{$a$-$b$} region, calculated their visibility
functions, and plotted the resulting values of $\zrec$ versus $\Delta z$.  Although we are not drawing
confidence regions in this $\zrec-\Delta z$ space, it can be clearly seen that the recombination
histories which are best able to reproduce observations lie in a cluster, with successively poorer histories lying further out.
As noted earlier,
for all of the ionization histories inside the $1\sigma$ region, 
$\zrec = 1055 \pm 25$, and $\Delta z
= 198 \pm 5$.

\begin{figure}[htbp]
\centering
\includegraphics{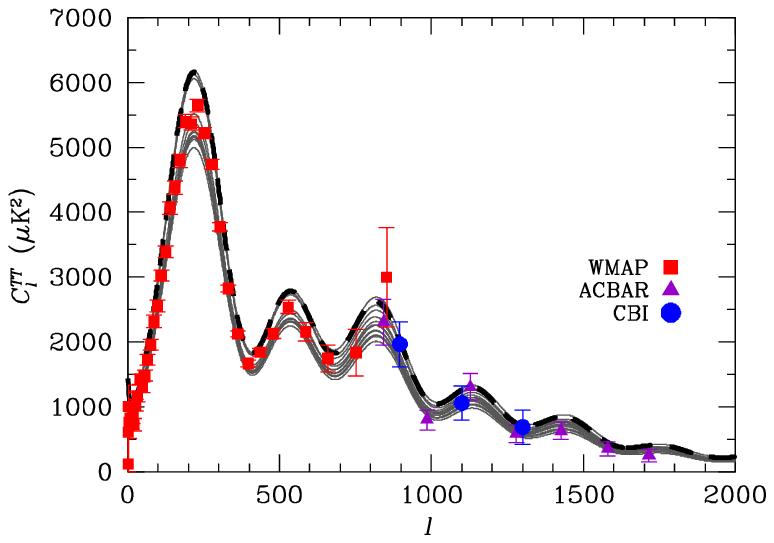}
\includegraphics{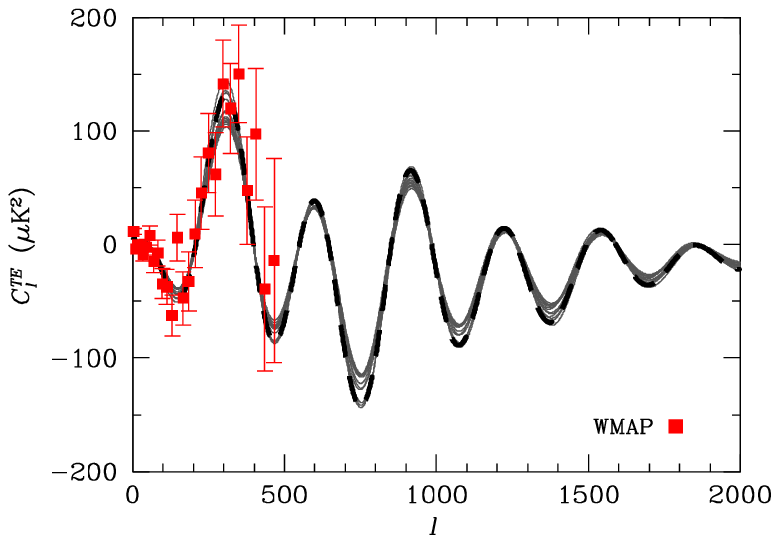}
\includegraphics{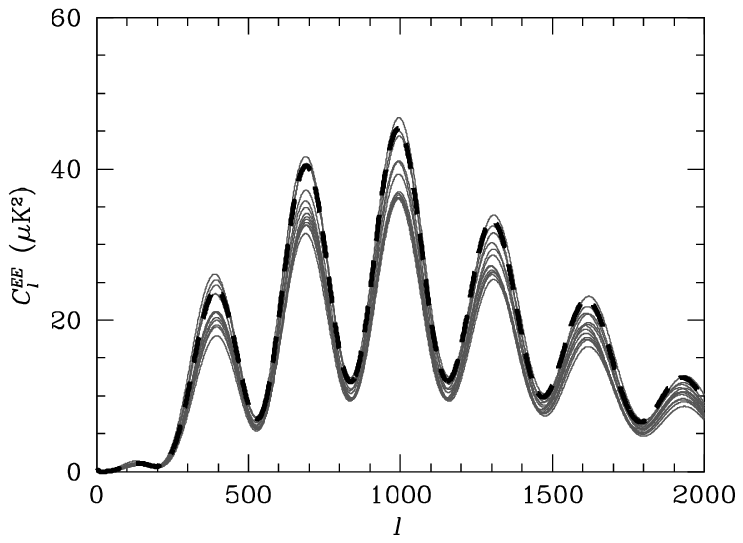}
\caption{\small $\Cl^{TT}$, $\Cl^{TE}$ and $\Cl^{EE}$ vs. $\ell$ for
recombination histories falling inside the $1\sigma$ region.  The thicker dashed line in each plot is standard recombination.}
\label{fig:cl_WMAP+_TTTE}
\end{figure} 
In Figure \ref{fig:cl_WMAP+_TTTE} we show the temperature, 
TE cross-correlation, and E-mode polarization power spectra 
for recombination histories falling within the $1\sigma$ region.  
The positions of the peaks in all three plots seem very well 
constrained, but the heights of the peaks differ somewhat.  
The $\Cl^{TT}$ and $\Cl^{TE}$ power spectra match the data 
very well, which is unsurprising since that was our criteria.  
Given the likely precision of data on the $\Cl^{EE}$ power 
spectrum in the near future, we conclude that it is unlikely 
that measurements of the E-mode polarization would be able to 
break any degeneracies in the allowed \hbox{$a$-$b$} parameter
space and significantly improve the constraints on recombination.

\section{Conclusions}  
\label{sec:Conclusions}
 
We have examined the new data from the WMAP mission, along with data from higher $\ell$ experiments, ACBAR and CBI, in an attempt to constrain recombination.
 We have included analyses of both the temperature power spectrum $\Cl^{TT}$ and the TE cross-correlation power spectrum $\Cl^{TE}$.   In the general parameterization that we used, the $1\sigma$ allowed region in \hbox{$a$-$b$} space was significantly contracted compared to earlier attempts with less precise data (\cite{Hannestad:2000}).

We find that the allowed region falls within $-0.6 < \loga < 0.5$ and $0.85 < b < 1.15$.
The average width and position of the surface of last scattering for recombination histories 
falling within the $1\sigma$ region are $\zrec = 1055 \pm 25$  and $\Delta z = 198 \pm 5$.
Note that the WMAP team estimated values for $\zrec$ and $\Delta z$, namely
\cite{Spergel:2003}
$\zrec^\text{WMAP} = 1088^{+1}_{-2}$ and $\Delta z^\text{WMAP} = 194 \pm 2$.
Our values for $\zrec$ and $\Delta z$ do not mean the same thing as those reported by the
WMAP team.  In the latter case, standard recombination was assumed, and these values simply give
the best-fit recombination history obtained by allowing the cosmological parameters to vary.  In our
case, we allow variations in the recombination history itself, so we expect a larger range of values
for $\zrec$ and $\Delta z$.  In fact, our values for $\zrec$  and $\Delta z$ are quite similar
to the WMAP values; this is a reflection 
of the fact that standard recombination is an 
excellent fit to the data and is approximately 
in the center of the distribution of the allowed 
variant recombination histories.
 
The changes in $a$
and $b$ are largely compensated by changes in other parameters.  The largest effects are that an increase in $a$ is matched by corresponding increases in
$\omega_b$ and $h$ and a corresponding decrease in $n_s$, while an increase in
$b$ is compensated by increases in $\omega_b$ and $h$. Tighter constraints
on these parameters, especially $h$, would allow us to narrow the range of allowed recombination histories.    
 
The result of these competing effects --- changes in $a$ and $b$ being partially compensated for by changes in the free cosmological parameters --- is that the anisotropy power spectra from the $1\sigma$ region all share very similar features.  In general, the positions of the peaks are strongly constrained, while the amplitudes of the peaks are less so.  The range in E-mode polarization power spectra corresponding to recombination histories from this region is very similar to the range in temperature and TE cross-correlation power spectra, and is unlikely to be useful in breaking degeneracies in \hbox{$a$-$b$} parameter space.
 
We find that the range of alternate recombination histories allowed by the WMAP, ACBAR, and CBI temperature and WMAP TE cross-correlation data is very small compared to that allowed by earlier data, and is approximately centered on standard recombination.  While there is still some room for non-standard recombination histories, the constraints have narrowed considerably with this new data.  Standard recombination fits the data extremely well, and there is no compelling
evidence from this study to suggest that a non-standard recombination is preferable.                                                    
             
\section{Acknowledgments}

A.M.L. and R.J.S. were supported in part by the Department
of Energy (DE-FG02-91ER40690) at Ohio State University.


\bibliographystyle{apj}

\end{document}